\documentclass[12pt]{aastex}
\usepackage{emulateapj5}
\usepackage{graphicx,psfig}

\def\ltsima{$\; \buildrel < \over \sim \;$}
\def\lsim{\lower.5ex\hbox{\ltsima}}
\def\gtsima{$\; \buildrel > \over \sim \;$}
\def\gsim{\lower.5ex\hbox{\gtsima}}

\newcommand{\be}{\begin{equation}}
\newcommand{\en}{\end{equation}}
\newcommand{\ergs}{\rm \ erg \; s^{-1}}

\def\cmdue {\rm \ cm^{-2}}

\def\deg {^\circ}

\begin{document}

\title{XMM-Newton observation of the 5.25 ms transient millisecond pulsar XTE
J1807--294 in outburst} 

\author{S.~Campana\altaffilmark{1}, M. Ravasio\altaffilmark{1},
G.L. Israel\altaffilmark{2,3}, V. Mangano\altaffilmark{2},
T. Belloni\altaffilmark{1}} 

\altaffiltext{1}{INAF-Osservatorio Astronomico di Brera, Via Bianchi 46, I--23807
Merate (Lc), Italy}

\altaffiltext{2}{INAF-Osservatorio Astronomico di Roma,
Via Frascati 33, I--00040 Monteporzio Catone (Roma), Italy}

\altaffiltext{3}{Affiliated to I.C.R.A.}


\begin{abstract}
We report on the results obtained for the millisecond transient X--ray pulsar
XTE J1807--294 in a 40 min orbital period system, based on an XMM-Newton ToO
observation carried out during March 2003. The source was found at a
luminosity level of about $2\times10^{36}\ergs$ in the 0.5--10 keV range
(assuming a distance of 8 kpc). We confirm the presence of the 5.25 ms
pulsations (after accounting for the orbital modulation) and find a
pulsed fraction of $5.8\%$ in the 0.3--10 keV band. The pulse 
shape in nearly sinusoidal.
The spectral continuum
of the source is well fitted by an absorbed Comptonization model plus a soft 
component. No emission or absorption lines have been detected in the 0.5--10
keV range with upper limits of 10--40 eV. The reported analysis represent the
first detailed study of this source, the fourth belonging to the
ultra--compact binary system class hosting an accreting neutron star. 
\end{abstract}


\section{Introduction}

In the past few years conclusive evidence that Low Mass X--ray Binaries
(LMXRBs) contain fast spinning, weakly magnetic neutron stars
has been gathered. Coherent periodicities have been discovered in four
sources SAX J1808.4--3654 (401 Hz, Wijnands \& van der Klis 1998),
XTE J1751--305 (435 Hz, Markwardt et al. 2002), XTE J0929--314 (185 Hz,
Galloway et al. 2002) and recently XTE J1807--294 (191 Hz,
Markwardt, Smith \& Swank 2003). 

All these four sources are transients, i.e. exhibit only sporadic activity 
and for most of the time remain in a state of low level activity 
(for a review see e.g. Campana et al. 1998a). 
These transient pulsars share many similarities  
among them, but are rather peculiar among transient sources containing a
neutron star (usually named Soft X--ray Transients, SXRT). Their peak X--ray
luminosities are rather faint ($\sim 0.01\,L_{\rm Edd}$), whereas SXRTs have
peak luminosities of $0.1-1\,L_{\rm Edd}$. 
They show X--ray pulsations, whereas for SXRTs and LMXRBs tight upper limits
exist (e.g. Vaughan et al. 1994). The discovery of X--ray 
pulsation during the outburst phase has also allowed to reveal an orbital
periodicity, which in all cases is shorted than $\sim 2$ hr (2.01 hr for SAX
J1808.4--3654, Chakrabarty \& Morgan 1998; 42.4 min for XTE J1751--305
Markwardt et al. 2002; 43.6 min for XTE J0929--314 Galloway et al. 2002; 40.1
min for XTE J1807--294, Markwardt, Juda \& Swank 2003). These periods are
much shorter than what usually found in SXRTs ($\gsim 4$ hr). These 
differences led some authors to suggest that transient pulsars might
form a separate (sub-)class among SXRTs, called faint transients (Heise et
al. 1998; King 2000; in't Zand 2001).

Faint transients are poorly known. The best studied source is SAX J1808.4--3658.
It was studied across three outbursts (Wijnands \& van der Klis 1998; Wijnands
et al. 2001), in quiescence (Campana et al. 2002), in the optical (Homer
et al. 2001) and radio (Gaensler et al. 1999) bands.

Concerning XTE J1807--294 very little is known. The source was discovered
during a Galactic plane scan by RXTE/PCA (Markwardt, Smith \& Swank 2003).
The source was observed at a level of 33, 38, 58, 41 and 20 mCrab (2--10 keV)
on 2003 Feb 16.7, 19.8, 21.6, 22.6 and Mar 13.9, respectively. A follow-up
Chandra observation on 2003 Mar 10.8 allowed to obtain a much better position
($1''$ uncertainty). Pulsations were still detectable and allowed a refined
orbital solution to be obtained from RXTE data (Markwardt, Juda \& Swank 2003).

Here we report on an XMM-Newton (discretionary time) TOO observation of 
XTE J1807--294 on 2003 Mar 22 (see also Campana et al. 2003).
In Section 2 we describe the observational data and the reduction strategy. 
In Section 3 we describe our results on spectral and timing analysis.
Conclusions are reported in Section 4.

\section{XMM-Newton observation and data analysis}

The XMM-Newton Observatory (Jansen et al. 2001) includes three 1500 cm$^2$
X--ray telescopes each with a European Photon Imaging Camera (EPIC, 0.1--15
keV) at the focus. Two of the EPIC imaging spectrometers use MOS CCDs (Turner
et al. 2001) and one uses pn CCDs (Str\"uder et al. 2001). Reflection Grating
Spectrometers (RGS, 0.35--2.5 keV, den Herder et al. 2001) are located behind
two of the telescopes. 

XTE J1807--294 was observed on March 22, 2003 from 13:57 to 18:40. MOS1 
was operating in Small Window mode, MOS2 in Prime Full Window mode and pn
in timing mode, all with thick filters. RGS1 and RGS2 were operating in the
Spectro + Q standard mode. Due to a problem in the field acquisition the 
source was originally pointed 1 arcmin apart. This problem was then corrected,
resulting in an overall efficiency decrease of the pn, MOS1 and RGS instruments.
We generated final product using {\tt sas} v. 5.4.1.
A small soft proton flare event occurred at the beginning of the observation, 
we excluded the first 2.5 ks from the analysis.
We obtained net exposure times of 14 ks and 9 ks for MOS2 and the
other instruments, respectively.

\subsection{MOS cameras}

The source is extremely bright such that the MOS cameras are saturated and the
radial support structure of the mirrors is visible.
Piled-up rate for the MOS2 camera is $\sim 8$ counts s$^{-1}$ (0.5--10 keV).
Source position is consistent with that determined with Chandra 
(Markwardt, Juda \& Swank 2003).

A simple way round to pile-up is to extract photons for spectral analysis in
an annulus around the source, thus excluding the inner (piled-up) core.
We followed this strategy for analysing the MOS2 data, whereas for MOS1 data the 
window is too small (see also the analysis carried out on Mkn 421, presenting
similar problems, by Ravasio et al. 2003, in preparation).
We extracted photons from an annulus of $30''$ and $60''$ inner and outer
radii, respectively. Background was extracted from two different outer 
circular regions $2'$ each. 
We selected only single-pixel events and make sure that no pile-up is present
(through the {\tt sas} tool {\tt epatplot}). This is also in line with
prescriptions by Molendi \& Sembay
(2003)\footnote{http://xmm.vilspa.esa.es/docs/documents/CAL-TN-0036-1-0.ps.gz}.
The count rate within the 0.5--10 keV energy band is $1.01\pm0.01$ counts
s$^{-1}$. Response matrix and ancillary files were then generated with {\tt
rmfgen} and {\tt arfgen}. Spectral data were rebinned to have 80 counts per
bin. 

\subsection{pn camera}

Timing mode was selected to have the opportunity to reveal X--ray pulsations. 
Given the high count rate of the source, this mode is particularly useful
to avoid any pile-up. We extracted the source spectrum from the raw image,
taking a column 10-pixel wide and the background from two separate 
boxes 5-pixels wide. We consider only ``single'' and ``double'' events.
No X--ray bursts were detected.  
The source count rate in the 0.5--10 keV band is $34.0\pm0.1$ counts s$^{-1}$. 
Background is at a level of $1.4\%$ in the full band.
Pile-up is negligible and dead time is at a level of $\sim 1.5\%$. Response
matrix and ancillary file for spectral analysis were generated using {\tt sas}
tasks. Spectral data were rebinned to have 80 counts per bin.

\subsection{RGS}

The standard {\tt sas} procedure {\tt rgsproc} was used to derive the first
order RGS spectra. 
In the 7--36 \AA\ band rates are $0.39\pm0.01$ and $0.56\pm0.01$ count
s$^{-1}$ for RGS1 and RGS2, respectively. For spectral analysis together with
the other instruments we rebinned the data to 100 counts per bin. 
We further inspect the RGS spectra for emission or absorption lines. In this
case we rebin the spectra to 20 counts per bin (see below). 

\section{Spectral analysis}

\begin{figure*}
\begin{center}
\psfig{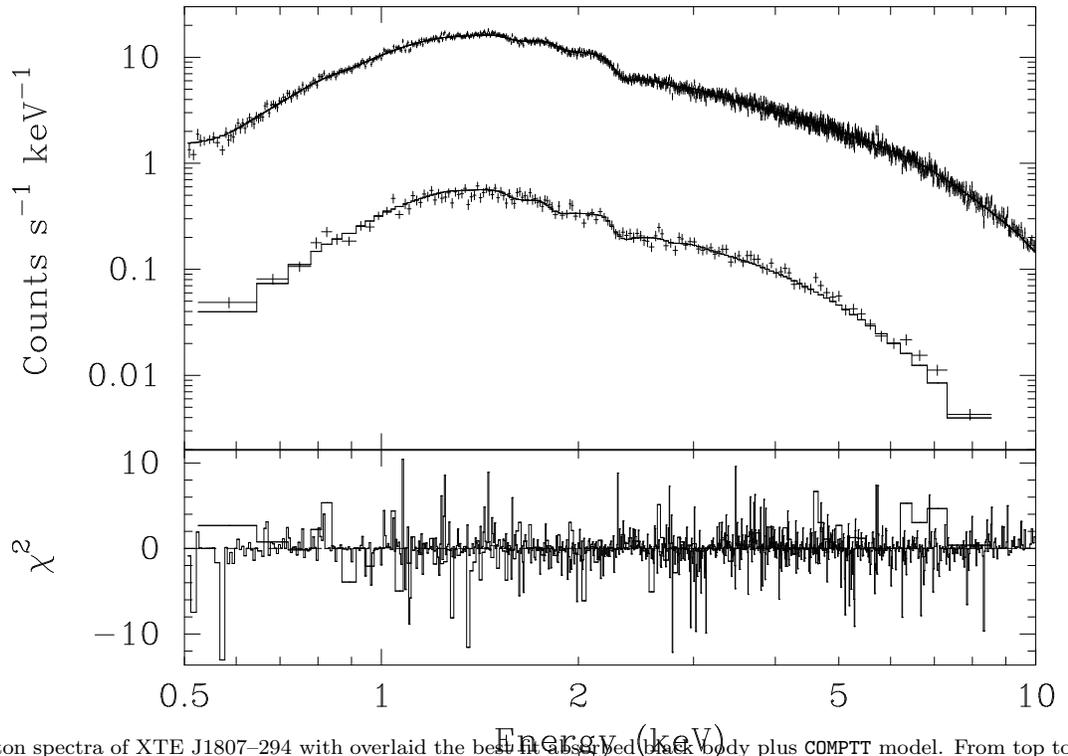}
\caption{XMM-Newton spectra of XTE J1807--294 with overlaid the best fit
absorbed black body plus {\tt COMPTT} model. From top to bottom there are the
pn and the MOS2 spectra. In the lower panel are shown the residuals in terms
of $\chi^2$.} 
\label{spe}
\end{center}
\end{figure*}

The spectral analysis has been carried out in the 0.5--10 keV energy range for
MOS2 and pn and in 7--36 \AA\ for the two RGSs, using XSPEC (v11.2). All
spectral uncertainties are given at $90\%$ confidence level for one degree of
freedom ($\Delta \chi^2=2.71$).  
We first tried the standard model for SXRTs, i.e. an absorbed soft component
(black body) plus a hard component (power law; e.g. Campana et al. 1998b). 
The fit is good with a $\chi^2_{\rm red}=1.10$ (1571 degrees of
freedom, d.o.f.). Fit parameters can be found in Table \ref{spetab}.
The column density is $(6.3\pm0.1)\times 10^{21}\cmdue$ which is slightly
higher than the Galactic value of $3\times 10^{21}\cmdue$. 
A comparable (but more physical) model is obtained by replacing
the power law with a Comptonization model like {\tt COMPTT} (Titarchuk 1994)
as observed in SAX J1808.4--3658 (Gierlinski, Done \& Barret 2002 but not in
XTE 1751--305, Miller et al. 2003). In this case we obtain  $\chi^2_{\rm
red}=1.10$ (1569 d.o.f., see Fig. \ref{spe}). 
Systematic errors at $1.6\%$ level provide an acceptable fit.
The column density is $4.6^{+0.3}_{-0.2}\times 10^{21}\cmdue$, consistent with
the Galactic value.
The black body temperature is $k\,T=0.80\pm0.03$ keV and its radius (assuming
spherical symmetry) is $R=(2.2\pm0.2)\,d_{8}^2\,f^2$ km (with $d_{8}$
the unknown distance to XTE J1807--294 in units of 8 kpc\footnote{The source
lies just in the direction of the Galactic center.} and $f$ the
spectral hardening factor obtained as the ratio of the color temperature to
the effective temperature, e.g. Merloni, Fabian \& Ross 2000). 
The temperature of the soft input photon is $0.24\pm0.02$ keV.
In the Comptonization model the temperature of the corona and the plasma
optical depth are tightly related. 
The best fit is for a relatively cool corona ($k\,T=9.6^{+102}_{-4.0}$ keV,
i.e. $k\,T>5.6$ keV) and an optical depth ($\tau<4.2$). 
A very hot corona ($k\,T>370$ keV) with a negligible optical depth ($\tau\sim
0.02$, $\tau<2.68$ at $90\%$ confidence level, c.l.) is also (formally)
acceptable. However, with a detector sensitive up to 10 keV, it is impossible
to test the presence of such a high coronal temperature.
The unabsorbed 0.5--10 keV flux for our best fit spectrum
amounts to $2.6\times 10^{-10}\ergs\cmdue$. 
Source luminosity is $2.0\times 10^{36}\,d_{8}^2\ergs$. The extrapolated
luminosity to the 0.01--100 keV band is a factor $\sim 1.7$ larger.

The hard component contributes almost the entire (unabsorbed) flux
($87\%$). The black body component is detectable only at low
energies. Other soft  components have been tried with similarly good results.
We cannot discriminate between an emission from a black body or from 
an accretion disk ($T_{d}\sim0.2$ keV, $R_{d}\sim 10$ km). A
neutron star atmosphere (G\"ansicke, Braje \& Romani 2002) and a thermal
spectrum however provided a worse fit to the data. 

\subsection{Line search}

We searched for an iron lines in the pn and MOS data with negative results.
Taking the line energy in 6.40--6.97 keV and 0.1 keV (best fit) or null values
for the line width, we derive an upper limit on the equivalent width of 18 and
25 eV, respectively. 

Following Miller et al. (2002) we searched the RGS spectra for emission or
absorption lines. We divided the spectra (rebinned at 20 counts per bin) in
slices of 3 \AA\ and inspect them for lines in the energy band 7--20 \AA. We
fitted the sliced spectra with an absorbed ({\tt VARTBABS}, Wilms, Allen 
\& McCray 2000) power law model.
All fits are acceptable. No emission or absorption lines are found. 
We directly searched for Ne K edge line at 14.25 \AA\ following the evidence for
enhanced absorption in ultra-compact binaries (Schulz et al. 2001; Juett,
Psaltis \& Chakrabarty 2001). No Ne edge is visible and the fit procedure
is not sensible to its presence. 
We also searched for Ne Ly$\alpha$ line at 12.13 \AA\ (in the range
11--13 \AA). No line is present and an upper limit of 41 eV can be set ($95\%$
c.l.). In case of a narrow line the limit is 9 eV. Detected Ne lines in 4U
1626--67 had equivalent witdhs of 20 \AA\ (Schulz et al. 2001).
  
\section{Timing analysis}

\begin{figure*}
\begin{center}
\psfig{figure=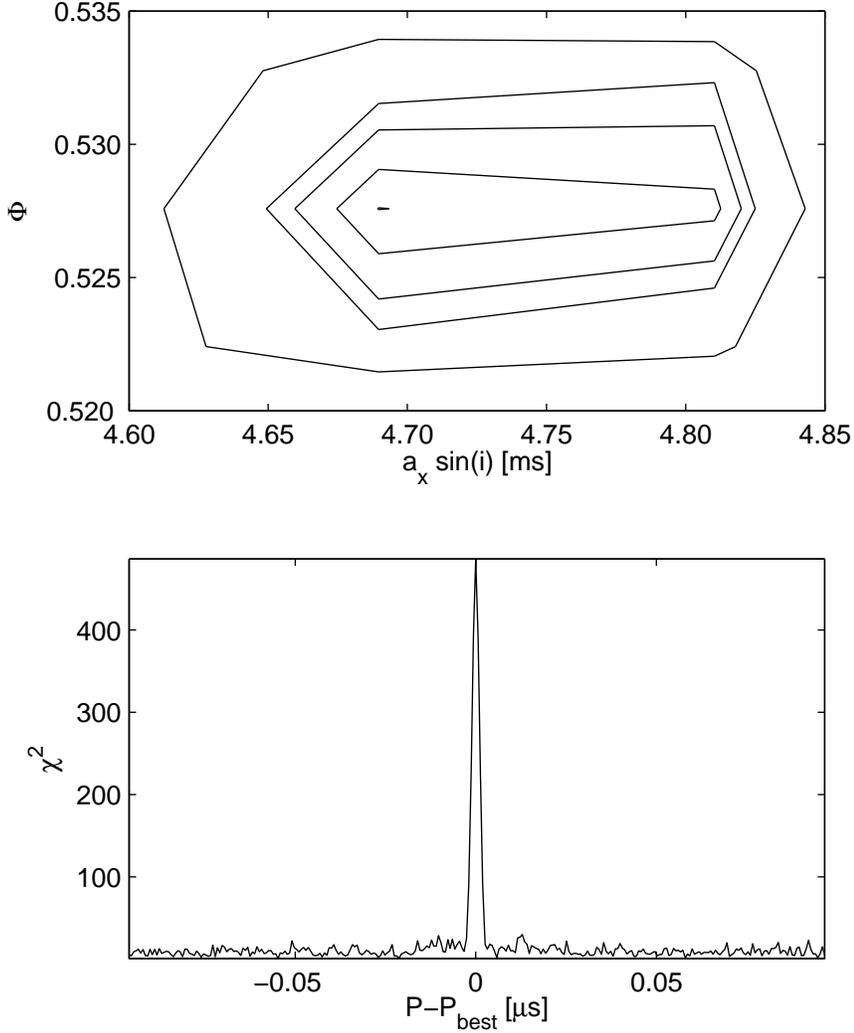,height=15cm}
\caption{
The upper panel shows the contour plot in the $a_x\,\sin i$ --
phase plane of the best correction for the orbital motion (assuming the
orbital period of 40.0741 min, Markwardt, Juda \& Swank 2003). Contours (from
outside) refers to $99.73\%$, $95\%$, $90\%$ and $68.3\%$ c.l.,
respectively (the best fit is marked with a small line). Orbital phases are in
arbitrary units (from the time of the first photon received by XMM-Newton).
In the lower panel the signal power (in terms of $\chi^2$) is shown. The
abscissa is in units of $\mu$s from the best period $P_{\rm best}=5.2459426$
ms. 
} 
\label{power}
\end{center}
\end{figure*}

\begin{figure*}
\begin{center}
\psfig{figure=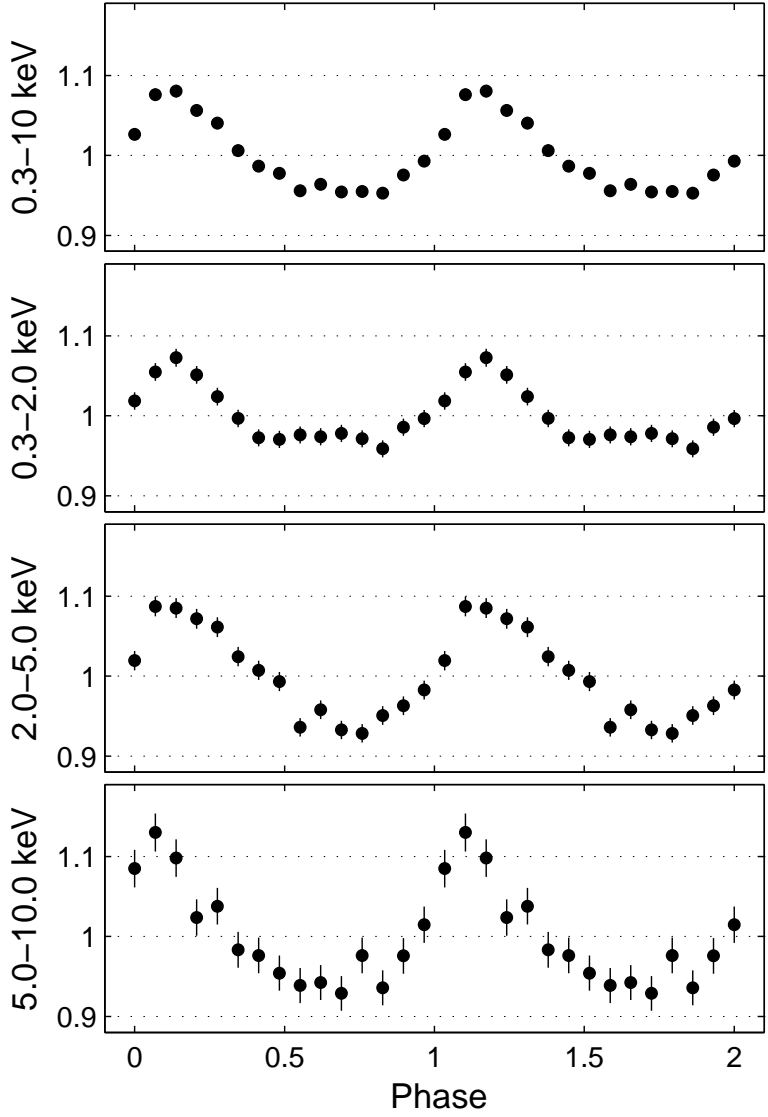,height=15cm}
\caption{Pn light curve of XTE J1807--294 folded at the neutron star spin
period. From top to bottom, these refer to 0.3--10 keV, 0.3--2 keV, 2--5 keV
and 5--10 keV, respectively. The curve is repeated twice for clarity.}
\label{contour}
\end{center}
\end{figure*}

We used the {\tt sas} task {\tt barycen} to correct the pn data. The time
resolution is 29.56 $\mu$s (Kuster et al. 2003).
A significant signal in the power spectrum is readily visible in the
data (Campana et al. 2003). However, the power spectrum shows a large number
of peaks likely resulting from aliasing of the short orbital period (see
Fig. \ref{power}). The correct pulse period can only be obtained after
correcting for the orbital modulation. We assumed the orbital period derived
by Markwardt, Juda \& Swank (2003) using RXTE data, i.e. $P_{\rm
orb}=40.0741$ min, and searched for the projected orbital radius and 
time of $90\deg$ mean longitude assuming the hypothesis of circular orbit. 
Stepping on those two parameters and, for each case, folding the events into a
10-bin light curve. The presence of a modulation in the folded light curve was
then estimated with a $\chi^2$ test. 
This is similar to the analysis carried out by Kirsch \& Kendziorra (2003).
We successfully recover the pulsation with a projected orbital radius
$a_x\,\sin i=4.7_{-0.04}^{+1.2}$ and an epoch of $90\deg$ mean longitude 
$T_0=52720.68644(13)$ ($90\%$ c.l., see 
Fig. \ref{contour}). Pulsations were detected at 5.2459426(2) ms ($90\%$ c.l.). 

After correction for the orbital motion and in the full band used for 
timing analysis (0.3--10 keV) the  (background subtracted) pulsed fraction amounts to 
$5.8\pm 0.3\%$ ($90\%$ c.l., without correction the pulsed fraction is only
$1.9\%$). The fit is not completely satisfactory ($\chi^2_{\rm red}=4.8$). An
additional sinusoidal component at half of the spin period is highly
significant (F-test probability $3.5\,\sigma$ and $\chi^2_{\rm red}=1.2$). The
pulsed fraction of XTE J1807--289 is comparable to what observed in SAX
J1808.4--3658 ($4\%$, Wijnands \& van der Klis 1998), XTE 1751--305 ($4\%$,
Markwardt et al. 2002) and XTE J0929--314 ($\sim 5\%$, Galloway et al. 2002).
The pulsed fraction increases with energy, going from $4.3\pm0.7$ in 0.3--2
keV to $7.4\pm0.7$ in 2--5 keV and to $7.5\pm1.4$ in 5--10 keV. Also for these 
curves two sinusoidal components provide a much better fit.

The orbital modulation is not directly visible in the data. 
Folding the barycentered pn light curve at the orbital period reported by 
Markwardt, Juda \& Swank (2003), we can put a $95\%$ upper limit on the 
modulated component of $1\%$.

%

\section{Conclusions}

We report on XMM-Newton observation of the fourth millisecond low mass X--ray
pulsar XTE J1807--294. We reveal the source still in a bright state 
$\sim 30$ d after its discovery. The source luminosity is $2\times 10^{36}\ergs$
at a fiducial distance of 8 kpc. 
 
Our results are very similar to what observed in XTE J1751--305 (Miller et al
2003). The hard spectral component contributes most of the observed flux
($87\%$), even though a soft component (a black body) is needed by the
data. The equivalent radius of the black body component is small,
$R=(2.2\pm0.2)\,d_{8}^2\,f^2$ km, suggesting an emitting hot spot onto the
neutron star surface. 

No emission or absorption lines are observed in the data. We
study in more detail Ne lines following the suggestion by Yungelson, Nelemans
\& van den Heuvel (2002) and Bildsten (2002) that the companions of faint
transient sources are Ne-rich white dwarves. We did not find any evidence of
emission lines (41 eV and 9 eV $95\%$ upper limits for a broad or narrow line)
or absorption edges.
Iron lines are not detected either with upper limits of $\sim 25$ eV.

The hard component (the main contributor to the source flux) can be better 
described with a Comptonization model rather than a simple power law model. 
This is at variance with the case of XTE J1751--305. The optical depth
and the plasma temperature are highly covariant. In any case a firm upper limit 
on the optical depth of 4.3 ($90\%$ c.l.) can be set. 
This is in line with the predictions of
Titarchuk, Cui \& Wood (2002). They suggested that X--ray pulsations can be
observed in sources with an optical depth lower than $\sim 4$ at a plasma
temperature of $\sim 20$ keV (as in our case). For sources with a higher optical
depth, e.g. $\tau=11.7\pm0.4$ in the Z source GX 349+2 (Di Salvo et al. 2001)
or $\tau=5.3\pm0.6$ in the atoll source 4U 1728--34 (Di Salvo et al. 2000),  
pulsations would be more easily suppressed by scattering.

\begin{acknowledgements}
We are grateful to the XMM-Newton project scientist, F. Jansen, for granting
time to this unsolicited target of opportunity observation. We thank the referee 
for her/his helpful comments. This work is partially supported through ASI and
Co-fin grants. 
\end{acknowledgements}

\clearpage

\begin{table*}[!htb]
\begin{center}
\caption{\label{spetab} Spectral fit with an absorbed black body component
plus a power law or a {\tt COMPTT} component.}
\begin{tabular}{cc}
\hline\hline
Parameter         & Value \\
\hline
$N_H$             & $6.3^{+0.1}_{-0.1}\times 10^{21}$ cm$^{-2}$\\
Black body $T$    & $0.81^{+0.04}_{-0.04}$ keV \\
Black body $R^*$  & $1.8^{+0.2}_{-0.2}$ km \\
Photon index      & $1.96^{+0.02}_{-0.02}$\\
$\chi^2_{\rm red}$& 1.10 (1569 d.o.f.)  \\ 
\hline
$N_H$             & $4.6^{+0.3}_{-0.2}\times 10^{21}$ cm$^{-2}$\\
Black body $T$    & $0.80^{+0.03}_{-0.03}$ keV \\
Black body $R^*$  & $2.2^{+0.2}_{-0.2}$ km \\
Input $T_0$       & $0.24^{+0.02}_{-0.02}$ keV \\
Plasma $T_p$      & $9.6^{+102}_{-4.0}$ keV \\
Optical depth     & $3.1^{+1.2}_{-2.1}$ \\
$\chi^2_{\rm red}$& 1.10 (1569 d.o.f.)  \\ 
\hline\hline
\end{tabular}
\end{center}
\tablecomments{$^*$ Black body radius at a distance of 8 kpc.}
\end{table*}

\end{document}